\documentclass[aps,pra]{revtex4}
\usepackage{graphicx}

\begin{document}

\title{Deterministic Secure Direct Bidirectional Communication Protocol \\
\thanks{*Email: zhangzj@wipm.ac.cn }}

\author{Zhanjun Zhang \\
{\normalsize Wuhan Institute of Physics and Mathematics, Chinese
Academy of Sciences, Wuhan 430071, China } \\
{\normalsize *Email: zhangzj@wipm.ac.cn }}

\date{\today}
\maketitle
\begin{minipage}{380pt}
In this letter a deterministic secure direct bidirectional
communication protocol is proposed by using the quantum
entanglement and local unitary operations on one photon of the
Einstein-Podolsky-Rosen (EPR) photon pair. \\

PACS Number(s): 03.67.Hk, 03.65.Ud\\
\end{minipage}

After the pioneering work of Bennett and Brassard published in
1984[1], a variety of quantum secret communication protocols have
been proposed( for a review see [2]). Majority of these protocols
are nondeterministic[1,3,4]. Only recently, several deterministic
secure direct communication protocols are proposed [5-7]. In all
these deterministic protocols, the secret message can only be
transmitted from one party to the other in a quantum channel,
i.e., two parties can not simultaneously transmitted their
different secret messages to each other in a quantum channel. In
general, convenient bidirectional simultaneous mutual
communications are very useful and usually expectant.
 Inspired by the deterministic secure direct protocol
(i.e., the ping-pong protocol) proposed by Bostr\"{o}m and
Felbinger [6], in this letter, by using the quantum entanglement
of the photon pair and two local unitary operations on one photon
in the pair by different parties in turn, we propose a
deterministic secure direct bidirectional simultaneous
communication protocol, which makes  secret timely mutual
communication (dialogue) possible in a quantum channel.

The ping-pong protocol allows the generation of a deterministic
key or even direct secret communication. A important idea of the
protocol is that the secure information is encoded by a local
operation on a photon of the Einstein-Podolsky-Rosen (EPR) photon
pair, which has already been raised by Bennett and Wiesner [8]. In
this letter, we will further take advantage of this idea to
complete the bidirectional communications. Our idea to use two
local operations performed by different parties to encode their
different bits is very important, because in terms of it the other
deterministic secure direct communication protocols using quantum
entanglement (e.g., Ref.[7]) can be easily changed into the
bidirectional ones besides the ping-pong protocol.

Let us start with the brief description of the ping-pong protocol.
Bob prepares two photons in the entangled state $|\Psi^+ \rangle =
(|0\rangle |1\rangle + |1\rangle |0\rangle ) / \sqrt{2}$ of the
polarization degrees of freedom. He stores one photon (home
photon) and sends the other one (travel photon) through a quantum
channel to Alice. After receiving the travel photon Alice randomly
switches between the control mode and the message mode. In the
control mode Alice measures the polarization of the travel photon
and announces the result publicly. After receiving Alice's result
Bob also switches to the control mode and measures the home photon
in the same basis and compares both measurement results, which
should be perfectly anticorrelated in the absence of Eve.
Therefore, the appearance of identical results is considered to be
the evidence of eavesdropping, and if it occurs the transmission
is aborted. In the other case, the transmission continues. In the
message mode, Alice performs the $Z_t^j  (j \in \{0,1\})$
operation on the travel photon to encode $j$ according to her
secret message and sends it back to Bob, where $Z=|0\rangle
\langle 0| - |1\rangle \langle 1|$. After receiving the travel
photon Bob measures the state of both photons in the Bell basis to
decode the $j=0(1)$ corresponding to the
$|\Psi^+\rangle(|\Psi^-\rangle= (|0\rangle |1\rangle -|1\rangle
|0\rangle ) / \sqrt{2}$) result. The security of the ping-pong
protocol has been proven [6].

Let us turn to our bidirectional communication protocol. We only
revise the ping-pong protocol in a subtle way. When Bob receives
the travel photon, he does not perform the Bell measurement on the
photon pair (i.e., the home photon and the travel photon) at once
but carry out a $Z_t^k (k \in \{0,1\})$ operation on the travel
photon (or the home photon) to encode $k$ according to his secret
message. Then he performs the Bell measurement on the photon pair
in his hand and publicly announces his measurement results. Since
Bob knows which local unitary operation he has performed on the
travel photon, according to his measurement result he can know
which local unitary operation Alice has performed on the travel
photon, i.e., he can extract Alice's encoding bit (See Table 1).
Similarly, since Alice knows which local unitary operation she has
performed on the travel photon, according to Bob's public
announcement she can know which local unitary operation Bob has
performed on the travel photon, i.e., she can extract Bob's
encoding bit (See table 1). Till now, a deterministic direct
bidirectional communication protocol has been proposed. \\

\begin{minipage}{370pt}
\begin{center}
Table 1.  Corresponding relations among Alice's, Bob's unitary
operations (i.e., the encoding bit) and Bob's Bell measurement
results on the photon pair. Alice's (Bob's)
unitary operations are listed in the first column (line). \\
\begin{tabular}{ccc}  \hline
              & $Z_t^0 (0)$           & $Z_t^1 (1)$
                            \\ \hline
$Z_t^0 (0)$  & $|\Psi^+ \rangle$& $|\Psi^- \rangle$ \\
$Z_t^1 (1)$  & $|\Psi^- \rangle$&$|\Psi^+ \rangle$ \\
\hline
\end{tabular} \\
\end{center}
\end{minipage}\\
\vskip 1cm

Let us discuss the security of our protocol. Before Bob's public
announcement, the present protocol is nearly same as the ping-pong
protocol, so it is secure for Bob to get the encoding bit from
Alice. Although Bob publicly announces his Bell measurement
result, because he has performed a unitary operation which Eve can
not know on the travel photon in his hand, also Eve can not know
which unitary operation Alice has ever performed on the travel
photon. Hence, it is secure for Bob to get the secret information
from Alice via our protocol. Now that Eve can not know which
unitary operation Alice has performed on the travel photon and Bob
publicly announces his measurement result on the photon pair,
Alice can securely know which unitary operation Bob has ever
performed on the travel photon. So it is also secure for Alice to
get the secret information from Bob.  Hence, the present
bidirectional communication protocol is secure against
eavesdropping. In fact, since the ping-pong protocol is secure
against the eavesdropping, it is obvious that the present protocol
is secure against eavesdropping. Incidentally, the ping-pong
protocol is not secure under Eve's intercept-measure-resent
attacks without eavesdropping [9]. The present protocol is also
not secure under Eve's intercept-measure-resent attacks without
eavesdropping. To fix this leak, we suggest that randomly both
Alice and Bob publicly announce corresponding fractions of their
unitary operations to check whether Eve is in the line. In
addition, the ping-pong protocol is insecure in a lossy channel,
as shown by W\'{o}jcik [10]. There is some probability that a part
of the secret message might be leaked to Eve. Similarly, the
present protocol is also insecure in the lossy channel. To fix
this leak, we still suggest that randomly both Alice and Bob
publicly announce corresponding fractions of their unitary
operations to check whether Eve is in the line.

In ping-pong protocol, the dense coding is abandoned in favor of a
secure transmission. By the way, it is said that the capacity of
the ping-pong protocol is doubled by introducing the dense coding
and the security is proven after the dense coding [11]. If so, an
improvement on the capacity of the present bidirectional
communication protocol is straightforward.

In conclusion, a deterministic secure direct bidirectional
communication protocol is proposed by using the quantum
entanglement and local unitary operations on one photon of the EPR
photon pair. As mentioned before, the subtle idea in this letter
to use two local operations performed by different parties to
encode their different bits is very important. In terms of the
subtle idea we are improving  other deterministic secure direct
communication protocols using quantum entanglement. We will show
them elsewhere.

This work is supported by the NNSF of China under Grant No. 10304022. \\

\noindent[1] C. H. Bennett and G. Brassard, in {\it Proceeding of
the IEEE International Conference on Computers, Systems, and
Signal Processing, Bangalore, 1984} (IEEE, New York,
1984),pp.175-179.

\noindent[2] N. Gisin, G. Ribordy, W. Tittel, and H. Zbinden, Rev.
Mod. Phys. {\bf 74},145 (2002).

\noindent[3] A. K. Ekert, Phys. Rev. Lett. {\bf 67}, 661 (1991).

\noindent[4] D. Bru$\beta$, Phys. Rev. Lett. {\bf 81}, 3018
(1998).

\noindent[5] A. Beige, B. G. Englert, C. Kurtsiefer, and H.
Weinfurter, Acta Phys. Pol. A {\bf 101}, 357 (2002).

\noindent[6] K. Bostrom and T. Felbinger, Phys. Rev. Lett.
{\bf89}, 187902 (2002).

\noindent[7] F. G. Deng, G. L. Long, and X. S. Liu, Phys. Rev. A
{\bf 68}, 042317 (2003).

\noindent[8] C. H. Bennett and S. J. Wiesner, Phys. Rev. Lett.
{\bf69}, 2881(1992).

\noindent[9] Q. Y. Cai, Phys. Rev. Lett. {\bf 91}, 109801 (2003).

\noindent[10] A. W\'{o}jcik, Phys. Rev. Lett. {\bf 90}, 157901
(2003).

\noindent[11] Q. Y. Cai and B. W. Li, {\it Imporving the capacity
of the Bostrom-Felbinger protocol}, accepted for publication in
Phys. Rev. A .

\end{document}